\documentclass[aps,prl,twocolumn,letterpaper,showpacs,showkeys]{revtex4}
\begin{document}
\title{A facet is not an island: step-step interactions and the 
fluctuations of the boundary of a crystal facet}
 \author{Alberto Pimpinelli$^\star$}
 \thanks{Corresponding author}
 \email[]{pimpinel@lasmea.univ-bpclermont.fr} 
 \author{M. Degawa}
  \author{T.L. Einstein}
  \email[]{einstein@umd.edu}
   \author{Ellen D. Williams}
\affiliation{Department of Physics, University of Maryland, College Park, Maryland 20742-4111
\\ $^\star$Also LASMEA, UMR 6602 CNRS/Universit\'e Blaise Pascal -- Clermont 2, 
F-63177 Aubi\`ere
cedex, France}

\begin{abstract}
\vspace*{2mm}
In a recent paper [Ferrari {\it et al.}, Phys.\ Rev.\ E {\bf 69}, 035102(R) (2004)],
the scaling law of the fluctuations of the step limiting a crystal facet has been
computed as a function of the facet size. Ferrari {\it et al.} use rigorous, but
physically rather obscure, arguments. Approaching the problem from a different
perspective, we rederive more transparently the scaling behavior of facet edge
fluctuations as a function of time.  Such behavior can be scrutinized with STM experiments and with numerical simulations.
\end{abstract}
\maketitle

In recent years it has become possible to probe quantitatively with STM the detailed atomic-scale fluctuations of steps near facet edges, most notably illustrated in extensive, painstaking measurements near (111) facets of Pb microcrystallites on a Ru support \cite{emundts,nowicki}.  Bonzel's review provides a thorough and lucid account of these investigations \cite{bonzel}.  Accordingly, it is an opportune time for theoretical examination of such systems.

In a very recent intriguing paper \cite{ferrari04}, Ferrari {\it et al.} have computed
the scaling of equilibrium fluctuations of an atomic ledge bordering a crystalline
facet. These authors find that the step-edge width $w$ scales as $w\sim L^{1/3}$
with the linear size $L$ of the facet. This result differs from what is expected,
and actually found, for the step bordering a 2D island, which performs a random
walk so that $w\sim L^{1/2}$. Ferrari {\it et al.} claim that the origin of the unusual
$L^{1/3}$ scaling lies in the step-step interactions between the facet ledge and
the neighboring steps.

Ferrari {\it et al.}'s calculation is based on the use of free fermions, transfer
matrix, random matrix properties, and specific models; it does not address the
question of the time behavior of step fluctuations.

In the present Communication, we approach the problem from the perspective of a
continuum-equation description of a faceted crystal.  The most easily
accessible experimental quantity is the step autocorrelation function
$G(t)= \langle [x(t)-x(0)]^2\rangle$, which is expected to have a power-law behavior at short
times: 
$G(t)\sim t^{2\beta}$. Hence, we compute the scaling of the ledge fluctuations with
time, using both the continuum-equation approach and the simple arguments developed
in Pimpinelli {\it et al.} \cite{pimpi}. 

Since the development in Ferrari {\it et al.} \cite{ferrari04} is rather obscure physically obscure, we readdress the problem from a
different point of view, based on continuum equations and scaling. Our approach
shows that the mean square width of a fluctuating crystal surface next to a facet
scales as $\langle(\delta z)^2 \rangle \sim \ell^{2/3}$, $\ell$ being the length scale in the
radial direction. We obtain then the time behavior of the surface fluctuations,
that will then be compared to that obtained from a different, more qualitative
approach (see below).

The projected free energy of a surface near a facet below its roughening
temperature is given by \cite{pivi,JW,pt}
\begin{equation}
f(\phi)=\gamma_0(T) + \frac{\beta(\theta,T)}{h}\tan \phi+g(\theta,T)\tan^3 \phi,
\end{equation}
where $h$ is the step height, $\gamma_0(T)$ is the terrace contribution, 
${\beta(\theta,T)}$ is the step
free energy per length, and $g$ the ``step interaction parameter" \cite{JW}. 
The angles $\phi$ and $\theta$ are the angle of the surface relative to
the facet and of the steps relative to an arbitrary direction, respectively.

The chemical potential for a crystallite in cylindrical coordinates $(r,\theta,z)$
reads \cite{md}
\begin{equation}
\label{2}
\mu(r)=\frac{\nu\beta(T)}{h r}+\frac{3\nu g(T)}{r}\left(\frac{dz(r)}{dr}
\right)^2+3\nu g(T)\frac{d}{dr}\left(\frac{dz(r)}{dr}
\right)^2,
\end{equation}
where the $\theta$ dependence of the step free energy and of the interactions has
been neglected for simplicity. Here $\nu$ is the
atomic volume, with $\nu/h$ then the atomic surface area.  It is straightforward to verify that the Pokrovsky-Talapov
equilibrium shape \cite{pt} is recovered from Eq.~(\ref{2}). In the present geometry, the
equilibrium shape is a stacking of circular layers separated by steps.

Consider the top step of the stack. Locally the ledge performs a random walk,
so that each time that the ledge moves forward or backward by one unit,
$r=\rho_0\pm\delta r$, the local surface height increases or decreases by one unit, 
respectively, $z(r)=z_0\pm \delta z$. (Note that the variable $\delta z$ is the continuum translation of $h$.)  In Eq.~(\ref{2}) the first term is the
Gibbs-Thomson contribution coming from the curvature of the layer, the second term
represents the variation in the interaction energy due to the change of the length
of the ledge when a single atom is removed or added. The last term represents the
change of interaction energy when the step-step distance in the curved part of the
crystal is varied. 

Thus, the contribution which dominates in Eq.~(\ref{2}) for small fluctuations of the
surface is the second one, since  the curvature and the step-step distance
are weakly affected by local step motions. Both the first and the last term in 
Eq.~(\ref{2}) can then be neglected. After next setting $1/r$ in the second term in 
Eq.~(\ref{2}) equal to $1/\rho_0$ to first order and letting $\partial (\delta
z)/\partial t=-K\nabla^2\mu +\eta_c$ (where $K$ is a transport coefficient, and
$\eta_c$ is a stochastic term describing mass-conserving random atomic motions at
the interface), we arrive at the equation
\begin{equation}
\label{c}
\frac{\partial (\delta z)}{\partial
t}=-K\nabla^2\left[\frac{3\nu g(T)}{\rho_0}\left(\frac{\partial(\delta
z)}{\partial r} \right)^2 \right]+\eta_c(r,t) .
\end{equation}

Eq.~(\ref{c}) looks like the nonlinear part of the so-called ``Montr\'eal'' or conserved KPZ model \cite{mont}. However, in the original model the
nonlinearity is absent in equilibrium, while here it stems from the contribution of
step-step interactions to the equilibrium chemical potential.

Since Eq.~(\ref{c}) is nonlinear, it cannot be solved analytically. Hence, we 
limit ourselves to a power-counting evaluation of the scaling exponents, following
Hentschel and Family \cite{hf}. Taking a characteristic length scale
$\ell$ as the unit, we can estimate the root-mean-square value of the time
derivative of the fluctuations as $\delta z_\ell/t_\ell$, and of their spatial
derivative as
$\delta z_\ell^2/\ell^4$. The root-mean-square value of the conserved noise term is
estimated as
$\eta_c\sim 1/(\ell^2S_\ell t_\ell)^{1/2}$, where $S_\ell=\sqrt{\ell^2+z_\ell^2}$
is the length of the fluctuating interface. For small-amplitude fluctuations, as 
is the case here, $S_\ell\approx \ell$, so that $\eta_c\sim 1/(\ell^3t_\ell)^{1/2}$
\cite{hf}. 

Equating the time derivative to the noise term yields
\begin{equation}
\label{tdnt}
\delta z_\ell\sim t_\ell^{1/2}/\ell^{3/2} .
\end{equation}

\noindent Equating the spatial derivative to the noise term yields
\begin{equation}
\label{sdnt}
\delta z_\ell^2\sim \ell^{5/2}/t_\ell^{1/2} .
\end{equation}

From Eqs.~(\ref{tdnt}) and (\ref{sdnt}) we finally get
\begin{equation}
\label{fea}
\delta z_\ell^3\sim \ell ,
\end{equation}
which has the same form as Ferrari {\it et al.}'s result. Note that the dynamics of step
fluctuations is also affected: From Eqs.~(\ref{tdnt}) and (\ref{fea}), e.g.,
we obtain
\begin{equation}
\label{time}
\delta z_\ell\sim t_\ell^{1/11} ,
\end{equation}
as well as
\begin{equation}
\label{timell}
\ell\sim t_\ell^{3/11} .
\end{equation}

Our argument allows extensions to nonconservative ledge fluctuations (i.e.\
fluctuations driven by attachment-detachment of atoms to and from the ledge). 
Indeed, if ledge fluctuations are driven by atom attachement-detachment, with
kinetic coefficient $\tilde K$, Eq.~(\ref{c}) has to be replaced by
\begin{equation}
\label{nc}
\frac{\partial (\delta z)}{\partial
t}=\tilde K\frac{3\nu g(T)}{\rho_0}\left(\frac{\partial(\delta
z)}{\partial r} \right)^2 +\eta(r,t) .
\end{equation}

Proceeding as before, we can estimate the time derivative of the fluctuations as
$\delta z_\ell/t_\ell$, their spatial derivative as $\delta z_\ell^2/\ell^2$, and
the (nonconserved) noise term as $1/(\ell t_\ell)^{1/2}$
\cite{hf}. In the latter relation we assume again small-amplitude ledge
fluctuations, consistently with the hindering action of neighboring steps.

Equating the time derivative to the noise term yields
\begin{equation}
\label{td}
\delta z_\ell\sim t_\ell^{1/2}/\ell^{1/2} .
\end{equation}

\noindent Equating the spatial derivative to the noise term yields
\begin{equation}
\label{sd}
\delta z_\ell^2\sim \ell^{3/2}/t_\ell^{1/2} .
\end{equation}

From Eqs.~(\ref{td}) and (\ref{sd}) we obtain 
\begin{equation}
\label{ncpl}
t_\ell\sim \ell^{5/3} ,
\end{equation}
as well as
\begin{equation}
\label{ncpl1}
\delta z_\ell\sim \ell^{1/3} .
\end{equation}
Together Eqs.~(\ref{ncpl}) and (\ref{ncpl1}) yield
\begin{equation}
\label{ncpl2}
\delta z_\ell\sim t_\ell^{1/5} ,
\end{equation}
as found above.

Note the Eq.~(\ref{nc}) looks like the KPZ equation without the linear term.
However, the nonlinearity stems here from equilibrium, small-amplitude fluctuations
of the step edge, and the resulting dynamics is different from that of the KPZ
model. The latter may be recovered from Eq.~(\ref{nc}) by assuming large-amplitude
fluctuations, so that $S_\ell\approx \delta z_\ell$, and $\eta_\ell\sim
1/(\delta z_\ell t_\ell)^{1/2}$.

Equating the time derivative to the noise term yields
\begin{equation}
\label{tdkpz}
\delta z_\ell^{3/2}\sim t_\ell^{1/2} .
\end{equation}

\noindent Equating the spatial derivative to the noise term yields
\begin{equation}
\label{sdkpz}
\delta z_\ell^{5/2}\sim \ell^2/t_\ell^{1/2} .
\end{equation}

From Eqs.~(\ref{tdkpz}) and (\ref{sdkpz}) we obtain 
\begin{equation}
\label{ncplkpz}
t_\ell\sim \ell^{3/2} ,
\end{equation}
as well as
\begin{equation}
\label{ncpl1kpz}
\delta z_\ell\sim \ell^{1/2} ,
\end{equation}
reproducing the power laws characteristic of the KPZ model. Note that in the limit of
unhindered, large amplitude fluctuations, the random walk scaling $w\sim
\ell^{1/2}$ is recovered, as expected.

Let us assume now that the results of Ferrari {\it et al.} hold, and let us use
Pimpinelli {\it et al.}'s argument \cite{pimpi} for finding the time scaling. The
argument computes the width
$\delta x$ of a fluctuation of linear size $\ell$ is found by letting the area of
the fluctuation,
$\ell \delta x$, equal to the fluctuation $\delta N$ of the number $N(t)$ of
particles involved in the mass-transport process responsible for the fluctuation,
during a time $t$. In turn,
$N(t)$ is estimated as 
\begin{equation}
\label{N}
N(t)\approx\frac {c_{eq}}{\tau^*}\ell L_s t,
\end{equation}
where ${c_{eq}}/{\tau^*}$ is the rate of the transport process, $c_{eq}$ being the
equilibrium particle density, and $\ell L_s$ the surface area feeding the
fluctuation
\cite{pimpi}. The assumption that mass transport is conservative, which is consistent
with Eq.~(3) above, yields ${c_{eq}}/{\tau^*}\approx c_{eq}D_e/\ell^2$, where $D_e$
is the edge diffusion coefficient. Also, $\ell L_s\approx a\ell$, $a$ being of the
order of the lattice spacing. Thus,
\begin{equation}
N(t)\approx tc_{eq}D_ea/\ell.
\end{equation}

Letting $\delta N= \sqrt N$, we now find
\begin{equation}
(\delta x)^2\ell^2\approx (\delta N)^2\approx tc_{eq}D_ea/\ell.
\end{equation}

Assuming the scaling relation $w\sim \ell^{1/3}$ yields finally
\begin{equation}
t\sim\ell^{11/3},
\end{equation}
or
\begin{equation}
\delta x(t)\approx t^{1/11},
\end{equation}
so that $G(t)\sim t^{2/11}$.

As in Ref. \cite{pimpi}, we can also compute the time behavior of a
fluctuation with non-conserved kinetics. This can be done by letting
${c_{eq}}/{\tau^*}\approx kc_{eq}$, with $k$ an appropriate kinetic coefficient for
atom detachment/attachment from and to the step edge. Then, using Eq.~(\ref{N}) with
$L_s\approx a$ yields
\begin{equation}
(\delta x)^2\ell^2\approx k {c_{eq}}\ell a t.
\end{equation}
Again letting $\delta x\sim \ell^{1/3}$ yields $\ell^{5/3}\sim t$, eventually
\begin{equation}
\delta x\sim t^{1/5}.
\end{equation}
Thus, the temporal scaling laws turn out to be identical to those computed above for
the fluctuations of the facet edge in the continuum equation approach.

In summary we have shown how the powerful general arguments used a decade ago to launch the systematic exploration of fluctuations of steps on vicinal surfaces can be extended to examine the fluctuations of steps near a crystalline facet.  We find rich and varied behavior that we hope will stimulate closer examination of fluctuation phenomena near step edges by both experiment and simulation.

\section*{Acknowledgment}  This work was supported by the MRSEC at University of Maryland 
under NSF Grant DMR 00-80008.

\end{document}